\documentclass{article}

\usepackage{PRIMEarxiv}
\usepackage{cite}
\usepackage[utf8]{inputenc} 
\usepackage[T1]{fontenc}    
\usepackage{hyperref}       
\usepackage{url}            
\usepackage{booktabs}       
\usepackage{amsfonts}       
\usepackage{nicefrac}       
\usepackage{microtype}      
\usepackage{lipsum}
\usepackage{tikz}
\usepackage{fancyvrb}
\usepackage{booktabs}
\usetikzlibrary{positioning, shapes.geometric, arrows.meta}
\usepackage{fancyhdr}       
\usepackage{graphicx}       
\usepackage{ulem}
\graphicspath{{media/}}     

\title{EEGEncoder: Advancing BCI with Transformer-Based Motor Imagery Classification
}

\author{
  Wangdan Liao\\
  Beihang University  \\
  \texttt{liaowangdan@buaa.edu.cn} \\
  \And
  Weidong Wang\\
   Chinese PLA General Hospital\\
  \texttt{wangwd301@126.com} \\
}

\begin{document}
\maketitle

\begin{abstract}
Brain-computer interfaces (BCIs) harness electroencephalographic signals for direct neural control of devices, offering a significant benefit for individuals with motor impairments. Traditional machine learning methods for EEG-based motor imagery (MI) classification encounter challenges such as manual feature extraction and susceptibility to noise.This paper introduces EEGEncoder, a deep learning framework that employs modified transformers and TCNs to surmount these limitations. We innovatively propose a fusion architecture, namely Dual-Stream Temporal-Spatial Block (DSTS), to capture temporal and spatial features, improving the accuracy of Motor Imagery classification task. Additionally, we use multiple parallel structures to enhance the model’s performance. When tested on the BCI Competition IV-2a dataset, our model results outperform current state-of-the-art techniques.
\end{abstract}


\keywords{motor imagery \and classification \and electroencephalography
(EEG) \and tansformer \and TCN \and Dual-Stream Temporal-Spatial Block (DSTS)}

\section{Introduction}
Brain-computer interfaces (BCIs) represent a cutting-edge technological frontier, offering a transformative approach to human-computer interaction. By facilitating direct neural communication, BCIs enable individuals to control external devices or systems through cerebral activity alone, bypassing conventional motor pathways.BCIs are particularly promising for applications in healthcare, rehabilitation, entertainment, and education. In the medical field, they provide a glimmer of hope for individuals with motor impairments, enabling the restoration of control over bodily functions. For example, BCIs have been instrumental in assisting individuals with spinal cord injuries to operate prosthetic limbs and have aided stroke survivors in regaining mobility\cite{ahmed2022artificial, altaheri2023deep}.

A critical BCI modality is EEG-based motor imagery (MI), which utilizes electroencephalographic (EEG) signals to deduce a user's intent for limb movement. MI signals, which are the brain's response to the mental rehearsal of motor actions, are essential for a BCI to identify the intended limb movement and control external devices accordingly.The intricate nature of MI-EEG signals, with their high-dimensional structure, calls for advanced machine learning and deep learning (DL) frameworks for effective interpretation. Traditional machine learning techniques, extensively used for MI-EEG signal classification, necessitate manual feature extraction and are susceptible to noise interference, often not fully capturing the signal complexity.

Recent endeavors in EEG motor imagery (MI) classification have started to harness the potential of transformer models, yielding encouraging outcomes\cite{jia2021mmcnn, altaheri2022physics, zhao2023eeg, li2020multi}. Despite these advancements, we posit that further enhancements are attainable. Our proposed model amalgamates the contextual processing prowess of transformers with the nuanced temporal dynamics captured by temporal convolutional networks (TCNs). This amalgamation is meticulously engineered to discern both the global and local dependencies that are characteristic of EEG signals. In our pursuit, we have also integrated cutting-edge developments from transformer architectures to bolster our model's efficacy. Our methodology represents a concerted effort to refine the interplay between transformers and TCNs, with the objective of bolstering the robustness and precision of EEG signal classification in a systematic and empirical fashion.

Our contribution: In this paper, we introduce EEGEncoder, a novel model for EEG-based MI classification that effectively combines the temporal dynamics captured by TCNs with the advanced attention mechanisms of Transformers. This integration is further augmented by incorporating recent technical enhancements in Transformer architectures. Moreover, we have developed a new parallel structure within EEGEncoder to bolster its robustness. Our work aims to provide a robust and efficient tool to the MI classification community, thereby facilitating progress in brain-computer interface technology.

\section{Related Work}\label{sec:related-work}
Motor Imagery (MI) classification has traditionally hinged on manual feature extraction methodologies, encompassing preprocessing, feature selection, and classification algorithms. Prominently, the Filter Bank Common Spatial Patterns (FBCSP) algorithm and its variants have been lauded for their efficacy in MI classification tasks \cite{ang2012filter}. Nevertheless, these conventional techniques often necessitate intricate preprocessing and elaborate feature engineering.

The emergence of Deep Learning (DL) has revolutionized MI classification by enabling direct learning of discriminative features from raw EEG data, thereby obviating the need for laborious manual feature extraction \cite{xu2020recognition, zhang2021hybrid, zhang2020motor}. Convolutional Neural Networks (CNNs) have emerged as a cornerstone in this domain, favored for their hierarchical feature extraction capabilities, which facilitate end-to-end learning. A plethora of CNN architectures, including Inception-CNN, Residual CNN, 3D-CNN, and multiscale CNNs, as well as those incorporating attention mechanisms, have been rigorously investigated \cite{jia2021mmcnn, lawhern2018eegnet, li2020multi, defferrard2016convolutional, amin2019deep}.

In parallel, Recurrent Neural Networks (RNNs) such as Long Short-Term Memory (LSTM) networks and Gated Recurrent Units (GRUs) have been employed to harness temporal dynamics within EEG signals \cite{luo2018exploring, tang2019motor}. However, Temporal Convolutional Networks (TCNs) have demonstrated their prowess over traditional RNNs in various sequence modeling tasks, showcasing their potential with an ability to expand the receptive field exponentially while maintaining a linear parameter increase and circumventing the vanishing or exploding gradient issues \cite{ingolfsson2020eeg, bai2018empirical, musallam2021electroencephalography}.

The advent of attention mechanisms, inspired by the human brain's ability to focus selectively on pertinent information, has further propelled advancements in EEG signal decoding. Since the inception of attention-based models by Bahdanau et al., such mechanisms have been extensively adopted in diverse domains, including natural language processing (NLP) and computer vision (CV) \cite{bahdanau2014neural, vaswani2017attention, kumar2021optical+}. Their integration into EEG signal processing has notably enhanced MI-EEG signal classification \cite{jia2021mmcnn, altaheri2022physics, zhao2023eeg, li2020multi, woo2018cbam, amin2021attention}.

To encapsulate, deep learning has substantially advanced the field of MI classification. Despite the strides made, the quest for optimal MI-EEG signal classification performance persists, inviting continuous exploration and innovation in the field.

\section{Method}

\begin{figure}[ht]
  \centering
  \includegraphics[width=0.9\textwidth]{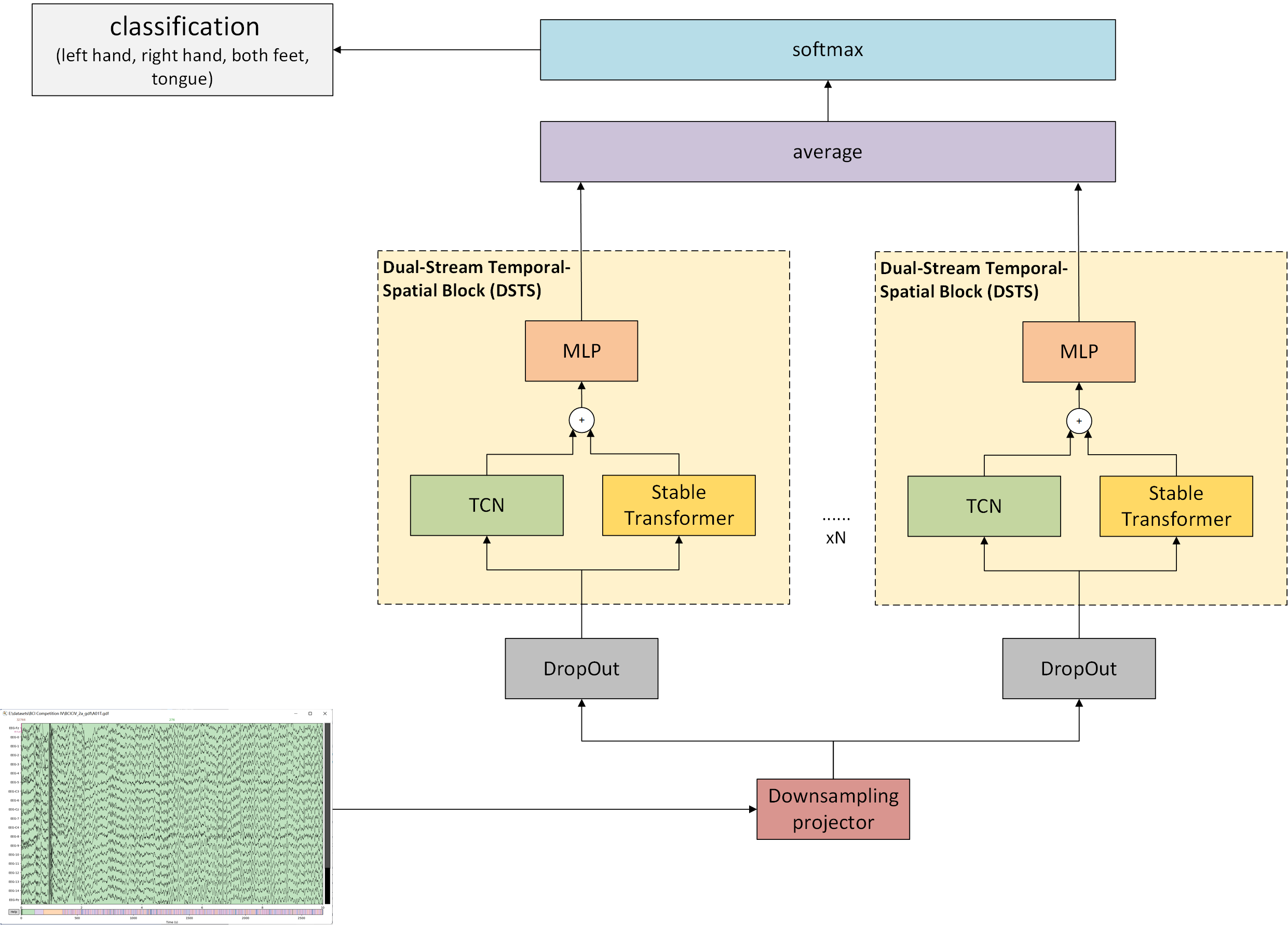}
  \caption{\textbf{Architecture of the EEGEncoder.} The figure illustrates the data processing pipeline within the EEGEncoder, highlighting the novel application of parallel dropout layers to enrich the diversity of the hidden state representations.}
  \label{fig:eeg_encoder}
\end{figure}

The overall architecture of EEGEncoder is depicted in Figure \ref{fig:eeg_encoder}. The proposed EEGEncoder model primarily consists of a Downsampling Projector and multiple parallel Dual-Stream Temporal-Spatial(DSTS) blocks, where the DSTS Block is formed by combining TCN and stable transformer. The Downsampling Projector preprocesses the MI-EEG signals through multiple layers of convolution. DSTS block integrates TCN and stable transformer to extract the temporal and spatial features of EEG signals. To enhance the robustness of the model, dropout layers are introduced before each parallel DSTS branch. The subsequent sections will provide a detailed description of the structure and function of each module.

\subsection{Downsampling projector for EEG Signal Preprocessing}

The Downsampling projector module within our EEG-based deep learning framework is designed to preprocess Motor Imagery EEG data, preparing it for intricate analysis by subsequent Transformer and Temporal Convolutional Network (TCN) layers. This module adeptly reshapes high-dimensional EEG sequences, characterized by a temporal resolution of 1125 and 22 channels, into a format that is conducive to convolutional processing. The main purpose of this process is to reduce the length of the sequence by passing continuous EEG signals through simple convolutional layers and average pooling layers. 

Considering the EEG data analogous to an image with dimensions $(1125, 22, 1)$, our approach involves the application of convolutional layers to extract spatial-temporal features, while concurrently mitigating noise and reducing inter-channel latency effects.

\begin{figure}
    \centering
    \includegraphics[width=0.3\linewidth]{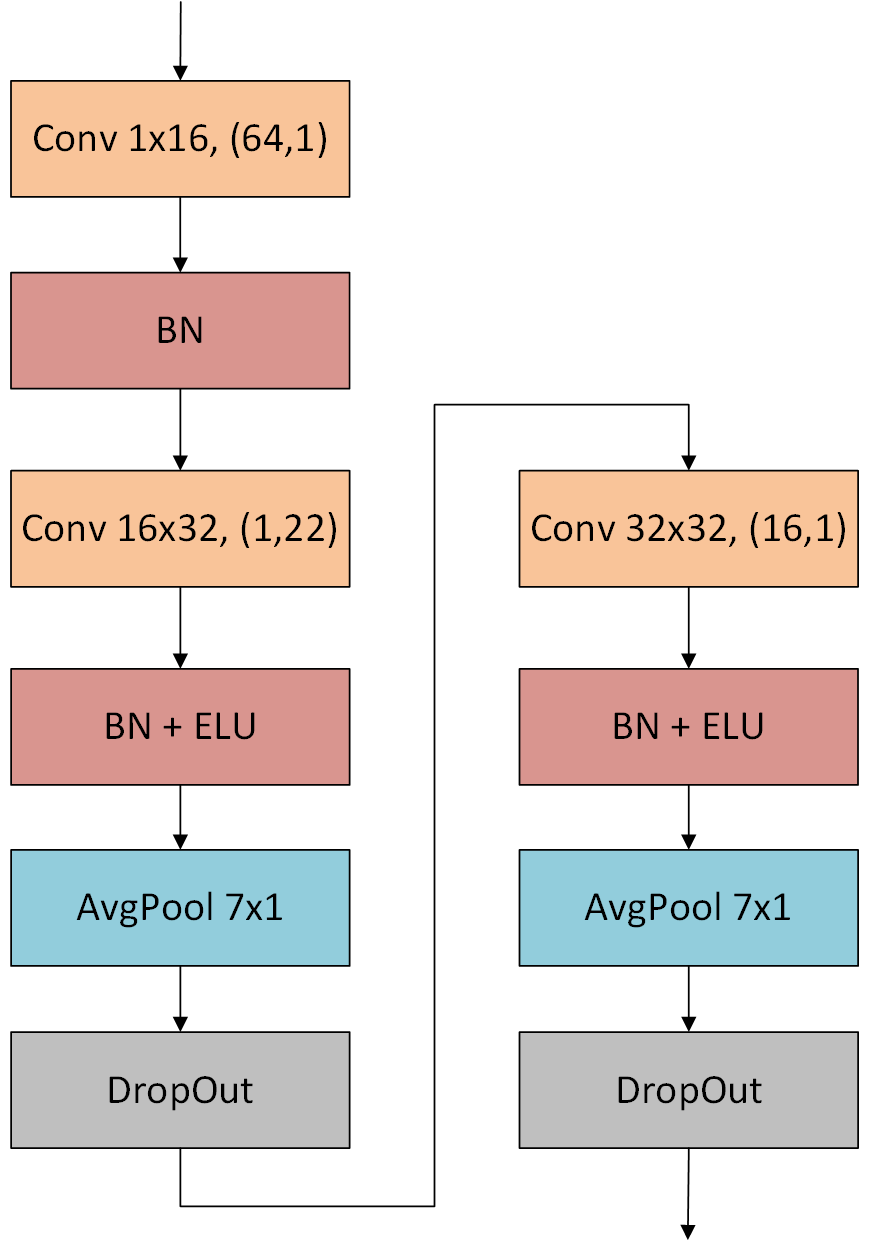}
    \caption{\textbf{Architecture of the Downsampling projector.} The figure provides a detailed schematic of the Downsampling projector's architecture. It includes three convolutional layers, with the second and third layers each followed by a batch normalization (BN) layer and an ELU activation layer. Additionally, two average pooling layers and two dropout layers are incorporated to foster model generalization. Specific parameters, such as the kernel size and stride for the convolutional layers, and the kernel size for the average pooling layers, are also depicted. For example, "Conv 1x16, (64,1)" signifies a convolutional layer transitioning from an input channel depth of 1 to an output channel depth of 16, with a stride of 64 along the width and 1 along the height of the input feature map.}
    \label{fig:downsampling projector}
\end{figure}

The core architecture of the Downsampling projector, as illustrated in Figure~\ref{fig:downsampling projector}, comprises three convolutional layers. The first convolutional layer is designed to initiate the feature extraction process without the application of an activation function. In contrast, the second and third convolutional layers are each followed by a batch normalization (BN) layer and an exponential linear unit (ELU) activation layer to stabilize the learning process and introduce non-linear dynamics into the model. The ELU \cite{Clevert2015Fast} activation function is defined as:

\begin{equation}
ELU(x) = 
\begin{cases} 
x & \text{if } x > 0 \\
\alpha(e^x - 1) & \text{if } x \leq 0 
\end{cases}
\end{equation}
where $\alpha$ is a hyperparameter that defines the value to which an ELU saturates for negative net inputs.

The second convolutional layer employs filters of size $(1, 22)$ to compress the channel dimension, effectively encoding channel-wise information into a singular spatial dimension. This strategic choice is informed by the understanding that variations among EEG channels are generally subtle and often predominantly due to noise.

Following this, average pooling layers with a stride of 7 are applied to reduce the temporal dimension. Interspersed dropout layers serve to promote regularization. 

\subsection{Stabilizing the Transformer Layer}

In the subsequent modules, we employ a modified Transformer layer \cite{touvron2023llama}, which has been adapted with recent technological advancements to enhance training stability and model efficacy. Here, we detail the specific alterations applied to the Transformer architecture.

Pre-normalization is a widely adopted strategy in deep learning, particularly for large-scale natural language processing (NLP) models like the Transformer. It is instrumental in stabilizing the training of very deep networks by addressing the vanishing and exploding gradient issues.

Unlike the standard Transformer architecture, where each sub-layer (such as self-attention and feed-forward layers) is succeeded by a residual connection and layer normalization (post-normalization), pre-normalization involves applying LayerNorm before each sub-layer.

Below is the simplified pseudocode for a Transformer block utilizing pre-normalization:

\begin{center}
\begin{BVerbatim}[baseline=c,boxwidth=auto]
def transformer_block_pre_norm(x):
    x = x + self_attention(norm(x))
    x = x + feed_forward(norm(x))
    return x
\end{BVerbatim}
\end{center}

The advantages of pre-normalization are manifold:

\begin{itemize}
\item \textbf{Enhanced Gradient Flow:} By normalizing inputs prior to each layer, we mitigate the risk of gradient vanishing or exploding during backpropagation, thus enabling the training of deeper architectures.
\item \textbf{Stable Training Dynamics:} Normalization ensures a consistent distribution of inputs across layers, fostering stability throughout the training phase.
\item \textbf{Quicker Convergence:} Pre-normalization has been associated with faster convergence rates in training models.
\end{itemize}

Our approach also incorporates RMSNorm, or Root Mean Square Layer Normalization \cite{zhang-sennrich-neurips19}, as the normalization function. RMSNorm diverges from traditional Layer Normalization by normalizing solely the standard deviation of activations, not the mean and standard deviation. It achieves this by dividing the activations by their root mean square, which maintains gradient scale and facilitates the training of deep networks.

\begin{equation}
\text{RMSNorm}(x) = \frac{x}{\sqrt{\frac{1}{N}\sum_{i=1}^{N}x_i^2 + \epsilon}}
\end{equation}

Here, $x$ represents the layer input, $N$ is the input's dimensionality, and $\epsilon$ is a small constant to prevent division by zero. This equation computes the RMS of the input and normalizes it by this value.

The key benefits of RMSNorm include:

\begin{itemize}
\item \textbf{Reduced Computational Burden:} RMSNorm obviates the need to compute the mean, thereby reducing computational demands relative to Layer Normalization.
\item \textbf{Stable Training:} By normalizing activation scales, RMSNorm aids in gradient flow, enhancing the overall stability of the training regimen.
\item \textbf{Compatibility with Deep Networks:} RMSNorm is particularly advantageous for deep networks, where it helps avert the typical gradient issues associated with such architectures.
\end{itemize}

To further enhance our model, we have substituted the typically-used ReLU activation with the Swish Gated Linear Unit (SwiGLU) \cite{shazeer2020glu}. SwiGLU is defined as the componentwise product of two linear transformations of the input, one of which is Swish-activated:

\begin{equation}
\text{SwiGLU}(x,W,V,b,c) = Swish_\beta (xW + b) \odot (xV + c)
\end{equation}

In the equation, \(x\) is the input, \(W\) and \(V\) are weight matrices, Swish  \cite{ramachandran2017searching} is defined as \(x \cdot \delta (\beta x)\), where \(\delta(z) = (1 + exp(-z))^{-1}\) is the sigmoid function and \(\beta\)
 is either a constant or a trainable parameter. SwiGLU's principal advantages are:
 
\begin{itemize}
\item \textbf{Computational Efficiency:} The gating mechanism of SWiGLU is notably efficient.
\item \textbf{Augmented Model Capacity:} It empowers the model to encapsulate more complex functionalities.
\item \textbf{Performance Enhancement:} SWiGLU typically boosts model performance across a range of tasks.
\end{itemize}

\subsection{Dual-Stream Temporal-Spatial Block}

\begin{figure}[ht]
  \centering
  \includegraphics[width=1\textwidth]{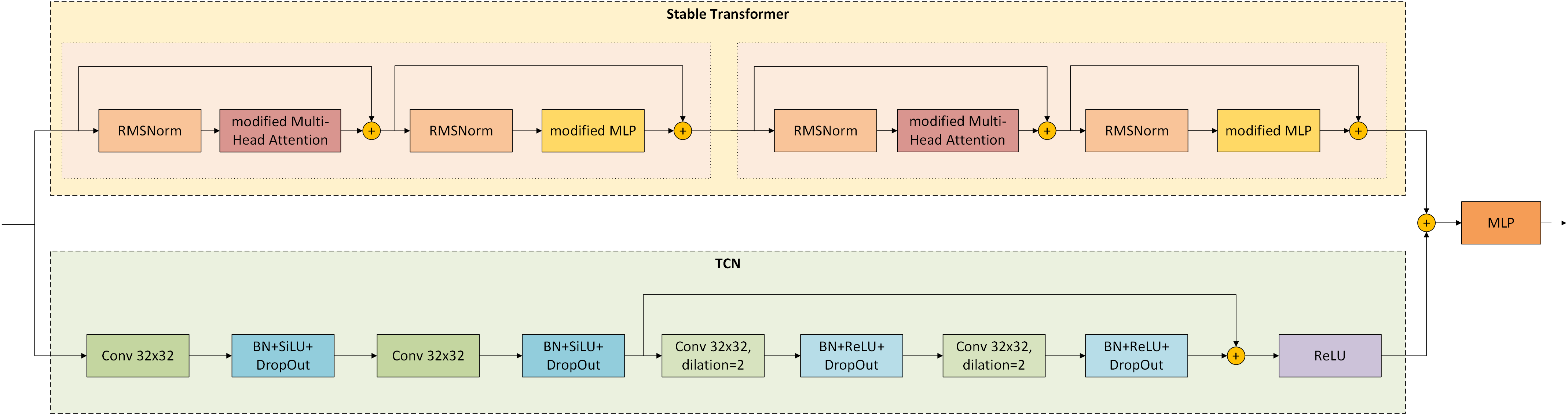}
  \caption{\textbf{Architecture of the DTDS Block.} The DTDS Block integrates a TCN for local temporal feature extraction with a self-attention block for global spatial context analysis, enabling a detailed examination of EEG signals for MI classification tasks.}
  \label{fig:dtds}
\end{figure}

The Dual-Stream Temporal-Spatial Block (DSTS Block) , as shown in Figure\ref{fig:dtds} presents an architecture specifically designed for the analysis of electroencephalogram (EEG) data during Motor Imagery (MI) tasks. This architecture integrates Temporal Convolutional Networks (TCNs) with stable Transformer modules, capitalizing on their complementary strengths to capture the temporal and spatial characteristics inherent in EEG signals.

TCNs utilize causal convolutions to process time-series data, effectively capturing temporal features with a high level of detail. The convolutional approach simplifies training and enhances feature extraction, which is particularly advantageous when dealing with the noisy and redundant nature of EEG data. However, the local focus of TCNs may result in insufficient representation of global dependencies, a notable challenge when analyzing extensive EEG sequences.

In contrast, Transformers employ a global self-attention mechanism that allows for the integration of contextual information across entire sequences. This capability enables the Transformer to perceive the broader context within the data, addressing a limitation of TCNs. Nonetheless, training Transformers can be complex, especially initially, and their performance may be less than optimal with the inherently noisy and complex EEG data. 

The DSTS Block is engineered to leverage the TCN's proficiency in local feature extraction and the Transformer's capacity for global context comprehension, thus aiming to provide a comprehensive analysis of EEG data. We also adopt the relative position representations as proposed by Shaw et al. in their seminal work\cite{shaw2018self}. This dual-stream approach is anticipated to improve the model's ability to identify patterns relevant to MI tasks by enhancing its analytical complexity. 

EEG data is processed through two distinct yet parallel pathways within the DTDS Block:

\begin{itemize}
\item The TCN pathway focuses on extracting local temporal features ($H_{\text{temporal}}$), utilizing causal convolutions to prioritize recent inputs and maintain temporal continuity.
\item The Transformer pathway is dedicated to identifying global spatial relationships ($H_{\text{spatial}}$), applying self-attention to consider inputs across the full sequence for a holistic spatial analysis.
\end{itemize}

To preserve the temporal sequence of EEG signals, a causal mask is integrated into the stable Transformer, ensuring information flow remains unidirectional. This approach is essential for maintaining the sequence's integrity, as it guarantees that predictions are based solely on past and present data:

\begin{equation}
H_{\text{temporal}}^{\prime} = \text{TCN}(H_{\text{temporal}})[:, -1, :]
\end{equation}
\begin{equation}
H_{\text{spatial}}^{\prime} = \text{StableTransformer}(H_{\text{spatial}}, \text{mask}=\text{causal})[:, -1, :]
\end{equation}

The variables $H_{\text{temporal}}^{\prime}$ and $H_{\text{spatial}}^{\prime}$ denote the final hidden states from the TCN and stable Transformer pathways, respectively, extracted from the last element in the sequence dimension. This selection strategy captures the accumulated temporal and spatial information up to the current moment.

These final hidden states are then integrated to create a composite feature representation, which is processed by a multi-layer perceptron (MLP) for the classification task:

\begin{equation}
H_{\text{integrated}}^{\prime} = H_{\text{temporal}}^{\prime} + H_{\text{spatial}}^{\prime}
\end{equation}
\begin{equation}
H_{\text{output}} = \text{MLP}(H_{\text{integrated}}^{\prime})
\end{equation}

The integration of TCN and Transformer pathways within the DTDS Block is designed to balance their respective strengths and limitations, enhancing the robustness and precision of BCI applications.

\subsection{EEG Signal Classification with EEGEncoder}

The EEGEncoder architecture represents a novel approach to the classification of electroencephalogram (EEG) signals. Traditional methodologies in this domain have frequently employed moving window techniques to extract temporal features from EEG data. These methods involve slicing the EEG sequence into overlapping temporal windows, which are then fed into the model to capture the dynamic aspects of the signal.

However, our architecture departs from this convention by harnessing the Transformer's intrinsic capability to contextualize data across the entire sequence. We postulate that this feature of the Transformer reduces the dependency on moving window slicing, thereby preserving the continuity and integrity of the temporal sequence.

To introduce variability and enhance the robustness of the model, we incorporate multiple parallel dropout layers. These layers independently introduce perturbations to the hidden states of the EEG sequence, a strategy designed to improve the model's performance by simulating a form of ensemble learning within the architecture itself.

After extensive experimentation and comparative analysis, we have optimized the EEGEncoder by configuring the DTDS block with a stable transformer consisting of four layers and two attention heads. Additionally, we have integrated five parallel branches, each comprising a dropout layer followed by a DTDS block. This configuration was determined to strike an optimal balance between model complexity and performance, leading to improvements in classification accuracy and generalizability.

\section{Results}

In this section, we provide a detailed evaluation of the EEGEncoder model, demonstrating its classification capabilities on the BCI Competition IV 2a dataset\cite{brunner2008bci}. We compare the performance of our model with various established models to underscore its effectiveness in decoding the complex patterns inherent in EEG signals for motor imagery tasks. The subsequent subsections elaborate on the model's performance metrics, a comparative analysis with other models, and discuss the significance of these results for the progression of brain-computer interface technologies.

\subsection{Datasets}

In our study, we primarily utilized the BCI Competition IV dataset 2a (BCI-2a) for training and evaluating the EEGEncoder model. The BCI-2a dataset comprises recordings from nine subjects across two sessions, with each subject performing 288 motor imagery trials. The trials were recorded using 22 EEG electrodes at a sampling rate of 250 Hz. The data underwent bandpass filtering between 0.5 and 100 Hz and a notch filter at 50 Hz to reduce power line interference.

The dataset encompasses four motor imagery (MI) tasks: left hand (class 1), right hand (class 2), feet (class 3), and tongue (class 4) movements. In our research, one session was used for model training, while the other was reserved for evaluation testing. The raw MI EEG signals from all bands and channels were fed into the model in the form of a \( C \times T \) two-dimensional matrix. Minimal preprocessing was applied to the raw data, employing a standard scaler to normalize the signals to have zero mean and unit variance.


Our research concentrates on the BCI-2a dataset due to its increased complexity and the greater challenge it presents, which better demonstrates the performance capabilities of our model.

\subsection{Training Configuration}

The model is trained with a specific set of parameters, as outlined in Table \ref{tab:training_config}. 

\begin{table}[ht]
\caption{Training Configuration for EEGEncoder.}
\centering
\begin{tabular}{lc}
\hline
\textbf{Config} & \textbf{Value} \\
\hline
Batch Size & 64 \\
Epochs & 500 \\
Optimizer & Adam \\
Learning Rate & 0.001\\
Loss Function & Cross Entropy \\
Label Smoothing & 0.1 \\
Dropout Ratio & 0.3 \\
Weight Decay (for MLP layers) & 0.5 \\
\hline
\end{tabular}

\label{tab:training_config}
\end{table}

The CrossEntropyLoss function is employed with label smoothing, which is set to a value of 0.1 to soften the target distributions, potentially improving the generalization of the model. To further regularize the training process and prevent overfitting, a dropout ratio of 0.3 is applied across the network, and weight decay with a coefficient of 0.5 is applied to all MLP layers.

\subsection{Results on BCIC IV 2a Dataset}
The EEGEncoder was subjected to a rigorous evaluation process on the BCI Competition IV dataset 2a. Each subject's data was split into two sessions, with one serving as the training set and the other as the test set. Accuracy and kappa score were selected as the primary metrics for assessment. To enhance the robustness of our findings, we performed five iterations of training and testing, each with a distinct random seed, and reported the average performance metrics.

Comparative analysis was conducted against three recent models known for their strong performance in the field: ATCNet, TCNetFusion, and EEGTCNet. The experiments were carried out using the implementations provided by Altaheri et al. \cite{altaheri2022physics}, which facilitated a consistent and fair comparison. The classification results for subjects 1 through 9 are detailed in Table~\ref{table:bcic2a_acc}.

The EEGEncoder demonstrated improved performance over the comparative models in eight of the nine subjects tested, with the exception being subject 4. The model exhibited particularly notable performance enhancements in subjects 2 and 5, suggesting an increased capability in managing the EEG signal variations observed in these individuals.

\begin{table}[ht]
\centering
\caption{Classification Performance for Subjects 1-9. Comparison of EEGEncoder, ACTCN, TCNetFusion, and EEGTCN Models in Terms of Accuracy (Acc) and Kappa Coefficient (Kappa) for Subjects 1-9}
\label{table:bcic2a_acc}
\begin{tabular}{lcccccccc}
\toprule
Subject & \multicolumn{2}{c}{EEGEncoder} & \multicolumn{2}{c}{ATCNet}\cite{altaheri2022physics} & \multicolumn{2}{c}{TCNetFusion}\cite{musallam2021electroencephalography} & \multicolumn{2}{c}{EEGTCNet}\cite{ingolfsson2020eeg} \\
\cmidrule(lr){2-3} \cmidrule(lr){4-5} \cmidrule(lr){6-7} \cmidrule(lr){8-9}
        & Acc(\%) & Kappa(\%) & Acc & Kappa & Acc & Kappa & Acc & Kappa \\
\midrule
1       & \textbf{86.46}&     \textbf{81.94}&       86.11&     81.5 &       79.17&     72.2 &       74.31&    65.7 \\
2       & \textbf{74.65}&     \textbf{66.2 }&       72.57&     63.4 &       64.24&     52.3 &       52.78&    37.0 \\
3       & \textbf{96.53}&     \textbf{95.37}&       93.06&     90.7 &       88.54&     84.7 &       88.89&    85.2 \\
4       & 81.94&     75.9&       \textbf{84.03}&     \textbf{78.7} &       64.93&     53.2 &       57.99&    44.0 \\
5       & \textbf{84.03}&     \textbf{78.7 }&       77.43&     69.9 &       71.53&     62.0 &       72.92&    63.9 \\
6       & \textbf{77.78}&     \textbf{70.37}&       73.61&     64.8 &       55.56&     40.7 &       43.75&    25.0 \\
7       & \textbf{95.83}&     \textbf{94.44}&       93.40&     91.2 &       86.81&     82.4 &       72.57&    63.4 \\
8       & \textbf{89.24}&     \textbf{85.65}&       86.81&     82.4 &       80.90&     74.5 &       77.43&    69.9 \\
9       & \textbf{91.67}&     \textbf{88.8}&       90.97&     88.3 &       80.21&     73.6 &       74.31&    65.7 \\
\bottomrule
\end{tabular}
\end{table}

The results indicate that the EEGEncoder not only excels in overall performance but also shows resilience in subjects where other models tend to falter. This resilience could be attributed to the model's architecture, which may be more adept at capturing the nuances of EEG signals across diverse cognitive tasks. However, further studies are warranted to confirm these findings and to explore the full potential of EEGEncoder in real-world BCI applications.

\subsection{Ablation Study}
To validate the efficacy of the various enhancements applied to the EEGEncoder, we conducted a series of ablation experiments. We began by consolidating the data from all nine subjects, merging their respective training and testing sets into single, comprehensive datasets. This approach allowed us to more effectively evaluate the generalizability of the model's improvements across different subjects.

Here, we present a selection of key experiments that were instrumental in assessing the impact of specific modifications. These experiments included removing the transformer component from the DTDS block, using 5 shift windows instead of five dropout branches, varying the number of transformer layers, adjusting the quantity of DTDS branches within the EEGEncoder, and comparing the performance of our modified stable transformer against the Vanilla Transformer. To ensure the statistical significance of our results, we averaged the outcomes across five iterations, each initialized with a different random seed. The summarized results are displayed in Table~\ref{tab:model_comparison}.

\begin{table}[h]
\caption{Performance Comparison of EEGEncoder With and Without Various Improvements.}
\centering
\begin{tabular}{lcc}
\toprule
 & Acc (\%) & Kappa (\%) \\
\midrule
\textbf{EEGEncoder (Full Model)} & 74.48 & 64.4 \\
Remove Transformer & 71.09 & 61.5 \\
shift window       & 74.43 & 64.2 \\
Transformer-2 Layer & 74.05 & 64.4 \\
Transformer-4 Layer & 74.04 & 64.2 \\
Transformer-8 Layer & 73.78 & 64.1 \\
1 DTDS Branch & 71.27 & 61.7 \\
3 DTDS Branch & 73.71 & 63.7 \\
7 DTDS Branch & 72.42 & 63.4 \\
9 DTDS Branches & 72.35 & 63.1 \\
Vanilla Transformer & 72.96 & 62.3 \\
\bottomrule
\end{tabular}
\label{tab:model_comparison}
\end{table}

The data in Table~\ref{tab:model_comparison} illustrates the impact of each modification on the EEGEncoder's performance. The removal of the transformer component led to a noticeable decrease in both accuracy and kappa score, underscoring its contribution to the model's effectiveness. Adjusting the number of transformer layers showed that a balance is needed to optimize performance, as evidenced by the slight decrease in accuracy with eight layers compared to two. Similarly, the number of DTDS branches was found to be a factor, with a single branch reducing performance and ten branches not improving it significantly. Lastly, the comparison between our stable transformer and the Vanilla Transformer variant indicates the importance of our modifications for achieving higher accuracy and kappa scores.

\section{Conclusion}

In our research, we have innovatively designed a model based on Temporal Convolutional Networks (TCN) and Transformers, specifically optimized for the classification of Motor Imagery (MI) signals derived from electroencephalograms (EEG). Our model introduces the DTDS block, a novel component that enhances the extraction of both local and global information from EEG data. By incorporating the Stable Transformer, we have stabilized the training process of the Transformer and reduced computational complexity. Furthermore, we have replaced the commonly used window shift technique with parallel multi-branch dropout+DTDS, which adds robustness and diversity to the feature extraction process.

The empirical evaluation of our proposed model has yielded promising results in the BCI Competition IV dataset 2a, where we achieved commendable performance without the need for complex preprocessing—relying only on a simple standard scalar. Looking ahead, our goal is to extend the training and validation of our model across a more diverse and extensive range of datasets. We aim to incorporate cutting-edge deep learning techniques, such as pre-training, to enhance the model's complexity and effectiveness. Ultimately, we aspire to achieve superior performance in MI classification tasks across a broader spectrum of categories, supported by larger datasets and more sophisticated model architectures.

\bibliographystyle{unsrt}  
\bibliography{references}

\end{document}